\documentclass[usenatbib,letter]{mnras}

\usepackage{newtxtext,newtxmath}

\usepackage[T1]{fontenc}
\usepackage{ae,aecompl}

\usepackage{graphicx}	
\usepackage{amsmath}	
\usepackage{amssymb}	
\usepackage{xspace}
\usepackage{xcolor}
\usepackage{aas_macros}
\usepackage{booktabs}
\usepackage[position=top]{subfig}
\usepackage{enumitem}
\usepackage[normalem]{ulem}

\usepackage{xpatch}

\usepackage{hyperref}
\usepackage{cleveref}

\makeatletter

\newcommand{\Mcc}{M_{\mathrm{cl},0}\xspace}
\newcommand{\Rcl}{R_{\mathrm{cl}}\xspace}
\newcommand{\Rshell}{R_{\mathrm{sh}}\xspace}

\newcommand{\Msun}{\mathrm{M}_{\odot}\xspace}
\newcommand{\Mso}{M_{*,1}\xspace}
\newcommand{\Mst}{M_{*,2}\xspace}

\title[Cloud recollapse in 30~Dor]{Forming clusters within clusters: How 30~Doradus recollapsed and gave birth again}

\author[Rahner et al.]{Daniel 
Rahner$^{1}$\thanks{daniel.rahner@uni-heidelberg.de},  Eric W.\ 
Pellegrini$^{1}$, Simon C. O. Glover$^{1}$, Ralf S. Klessen$^{1,2}$ \\
$^{1}$ Universit{\"a}t Heidelberg, Zentrum f{\"u}r Astronomie, Institut f{\"u}r Theoretische Astrophysik, \\
Albert-Ueberle-Stra{\ss}e 2, 69120 Heidelberg, Germany\\
$^{2}$ Universit{\"a}t Heidelberg, Interdisziplin{\"a}res Zentrum f{\"u}r Wissenschaftliches Rechnen,\\
Im Neuenheimer Feld 205, 69120 Heidelberg, Germany}

\date{Accepted 20/09/2017. Received 19/09/2017 ; in original form 13/09/2017}

\pubyear{2017}
\begin{document}

\label{firstpage}
\pagerange{\pageref{firstpage}--\pageref{lastpage}}
\maketitle

\begin{abstract}
The 30 Doradus Nebula in the Large Magellanic Cloud (LMC) contains the massive starburst cluster NGC~2070 with a massive and probably younger stellar sub clump at its center: R136. It is not clear how such a massive inner cluster could form several million years after the older stars in NGC~2070, given that stellar feedback is usually thought to expel gas and inhibit further star formation. Using the recently developed 1D feedback scheme \textsc{warpfield} to scan a large range of cloud and cluster properties, we show that an age offset of several million years between the stellar populations is in fact to be expected given the interplay between feedback and gravity in a giant molecular cloud (GMC) with a density $\gtrsim 500$\,cm$^{-3}$ due to re-accretion of gas onto the older stellar population. Neither capture of field stars nor gas retention inside the cluster have to be invoked in order to explain the observed age offset in NGC~2070 as well as the structure of the interstellar medium (ISM) around it.
\end{abstract}


\section{Introduction}
30 Doradus (a.\,k.\,a.\ the Tarantula Nebula) is a massive star-forming region in the Large Magellanic Cloud (LMC) with a complex geometry. In it resides the massive ionizing star cluster NGC~2070 which hosts the compact subcluster R136 (formally known as RMC~136) at its core. Numerous studies have concluded that R136 is a distinct stellar population considerably younger ($\sim$1\,Myr) than the other stars in NGC~2070 ($\sim$5\,Myr). This gives rise to the question of how two distinct stellar populations could form there.

Multiple generations of stars are usually discussed in the context of globular clusters, and various mechanisms for the formation of younger generations have been suggested, such as recollapse of the  ejecta of first generation asymptotic giant branch stars \citep{Ercole2008}, fast-rotating massive stars \citep{Decressin2007}, or interacting massive binaries \citep{DeMink2009}.
However, this debris accounts for only a small fraction of the mass of the first generation of stars. It is therefore difficult to form a massive second population in this fashion, particularly if the age separation between the populations is small, as in the case of NGC~2070. Furthermore, bimodal (or even multimodal) age distributions are not limited to globular clusters. They are observed in young star clusters, too, with age separations ranging from tens of Myr in Sandage-96 and possibly NGC~346 \citep{Vinko2009, DeMarchi2011a} 
to 1\,Myr or less in the Orion Nebula Cluster \citep{Beccari2017}.

On the one hand, stellar feedback is often assumed to disrupt molecular clouds and prevent or at least reduce further star formation \citep{Murray2011, Wang2010}. On the other hand, feedback can also be positive, compressing the interstellar medium (ISM) into dense shells and triggering star star formation \textit{around} the first generation star cluster \citep{Koenig2012}. However, neither outcome is consistent with what is observed in the 30 Doradus region: The existence of a massive young cluster \textit{within} another massive cluster produces serious challenges to normal models of star formation and feedback.

Thus, the old stellar population in NGC~2070 must not only have failed to destroy its natal cloud, but the cloud must also have retained or re-accreted enough dense gas to successfully form another massive cluster in a second burst of star formation.
\citet{Silich2017} showed that in dense conditions (e.\,g. a clump with $n \gtrsim 10^5$\,cm$^{-3}$ in a $10^6\,\Msun$ cloud) stellar winds may not be strong enough to clear all of the gas out of a newly-formed star cluster. Retained gas could then cool and form a second generation of stars inside the old star cluster, explaining how multiple stellar populations can form at the same location. However, the authors do not predict when the second formation of stars might form.

As we will show, even if no gas remains in the cluster, a second generation of stars might still form from gas that has been expelled but is later re-accreted. \citet{Pflamm-Altenburg2009} presented a static model which explains accretion onto a massive star cluster, but their model is only applicable to clusters with stellar masses above $10^6\,\Msun$ and thus is not applicable in the case of NGC~2070.
In a recent paper \citep{Rahner2017a} we presented a 1D dynamical model which accounts correctly for all major sources of feedback in isolated, massive star-forming regions. 
A major result of that work was the prediction of a relatively uniform recollapse time for molecular clouds where feedback is insufficient to disrupt the molecular cloud. 
In this paper, we apply our model to the cloud responsible for forming NGC~2070 and R136 and show that it provides a viable explanation for the origin of the double cluster in 30 Doradus, as well as explaining the bulk of the observed nebular structure. 
The low-shear environment of the LMC \citep{Thilliez2014} renders 30~Dor a good test case for the model.

\section{Model}
The morphology of 30~Dor is the superposition of numerous shell-like structures, seen in multiple phases of the ISM from ionized, neutral and molecular emission. The ionized gas forms bubbles containing hot, X-ray emitting gas \citep{Townsley2006}. Using [SII]/H$\alpha$ observations, \citet{Pellegrini2011} showed that the H\,{\sevensize\bf II} region around NGC~2070 has the shape of hemispherical bowl with a radius of $40 - 60$\,pc. R136 is offset approximately 12\,pc from the center of the bowl. The distance between R136 and the shell is thus $\sim 30-70$\,pc.
 
There is no consensus about the exact mass and age of the different populations in NGC~2070. 
Most observations agree, however, that there is an older and a younger population in NGC~2070\footnote{Some, e.g. \citet{Selman1999}, even report on three distinct starbursts. However, the age of the intermediate population might be the result of misclassification of main-sequence Wolf-Rayet stars and we will focus on just two populations in this work.}. The older stars have an age of $\sim 3-7$\,Myr \citep{Brandl1996,Walborn1997,Selman1999,Sabbi2012,Cignoni2015}, while the younger population, which is mainly concentrated in R136, is $\sim 0.5-2$\,Myr old \citep{Massey1998,Selman1999,Sabbi2012,Crowther2016}. The mass of R136 is $2.2\times 10^4 - 1\times 10^5\,\Msun$ \citep{Hunter1995,Andersen2009,Cignoni2015} compared to NGC~2070's mass as a whole of $6.8\times 10^4 - 5\times 10^5\,\Msun$ \citep{Selman1999,Bosch2001,Bosch2009,Cignoni2015}.  At a projected distance of $\sim$40\,pc from R136 lies the much older star cluster Hodge 301. Due to its comparatively low stellar mass ($\sim 6000\,\Msun$, \citealt{Grebel2000}) and its much weaker feedback we will ignore it here. The gas in the 30~Dor nebula has a mass of the order $10^6\,\Msun$ \citep{Dobashi2008, Faulkner1967} to $10^7\,\Msun$ (as estimated by \citealt{Sokal2015} from H\,{\sevensize\bf I} measurements in \citealt{Kim2003}).

 Starting from these observations we propose the following timeline to explain the structure of the inner 30~Dor region (shown in Figure~\ref{fig:expansion_example}):
\begin{itemize}
\item A massive cluster (the old population in NGC~2070) forms in a giant molecular cloud (GMC), the properties of which are rather unconstrained. Stellar feedback compresses the surrounding gas into a thin shell and accelerates it outwards.
\item Feedback is not strong enough to unbind a significant fraction of gas from the cloud. Instead, the swept-up material stalls, possibly breaks up into fragments, 
and recollapses towards the cluster under the influence of gravity.
\item As the gas collapses back onto NGC~2070, compressing the gas up to very high densities, a new, young, massive star cluster forms at the centre of NGC~2070: R136.
\item The combined feedback from R136 and the old population in NGC~2070 leads to a renewed expansion, giving rise to the shell around NGC~2070 as it is observed today.
\end{itemize}

We use the 1D feedback code \textsc{warpfield} \citep{Rahner2017a} to model the expansion of the shell around NGC~2070 and to test the hypothesis described above. While we acknowledge that the latest 3D simulations of feedback in molecular clouds also include a wide range of relevant physics \citep[e.\,g.][]{Wareing2017,Peters2017,Geen2016}, it is still prohibitively expensive to run a high number of simulations probing a large parameter space of initial conditions. Although some 3D results are not reproducible in 1D codes and could provide alternative explanations of the double cluster in 30~Dor, the employed 1D model will help constrain the initial conditions that should be explored in computationally more demanding studies of 30~Dor progenitors (Rahner et al. in prep).

\textsc{warpfield}, which is derived in part from observational studies of feedback in nearby star-forming regions, accounts for feedback from stellar winds, supernovae (SNe), direct and indirect radiation pressure, and gravity (self-gravity of the swept-up shell and gravity between the shell and the stars). We treat cooling of the hot, X-ray emitting wind bubble in an approximate fashion as in \cite{MacLow1988}. The photoionized and neutral gas are set to constant temperatures of $10^4$ and $10^2$~K, respectively.
We assume a metallicity of $Z=0.43\,Z_{\odot}$ for the gas and the stars in 30~Dor \citep{Choudhury2016}. Stellar evolution is modeled with \textsc{starburst99} \citep{Leitherer1999,Leitherer2014} using Geneva evolution tracks for rotating stars up to 120\,$\Msun$ by interpolating between models by \citet{Ekstrom2012} and \citet{Georgy2012}. As there is evidence that the initial mass function (IMF) in 30~Dor extends to $300\,\Msun$ \citep{Crowther2010,Khorrami2017a}, for the first starburst we set the upper mass limit of the assumed \citet{Kroupa2001} IMF to $300\,\Msun$. The most massive stars in R136 still exist and their masses have been determined \citep{Crowther2010}, so for the second starburst, we use a \citeauthor{Kroupa2001} IMF up to 120\,$\Msun$ and add 4 additional stars of 165, 220, 240, and $300\,\Msun$ to the cluster. To estimate the mass-loss rate $\dot{M}_{\rm{w}}$, bolometric luminosity $L_{\rm{bol}}$, and emission rate of ionizing photons $Q_{\rm{i}}$ for stars in the range $120 - 300\,\Msun$ on the main sequence, we interpolate between models presented in \citet{Crowther2010}. The terminal wind velocity of these stars is set to 2500\,km\,s$^{-1}$ and the ionizing luminosity $L_{\rm{i}} = 0.6 L_{\rm{bol}}$, as calculated with \textsc{wmbasic} \citep{Pauldrach2001} for a 50\,000\,K star. We limit SN feedback to zero-age main-sequence masses between 8 and 40\,$\Msun$ because more massive stars at subsolar metallicity are expected to end their lives in weak SN~Ib/c explosions or as direct-collapse black holes \citep{Heger2003}. However, our results are not strongly affected by the choice of the upper mass cut-off. 

  Since the actual cloud and cluster parameters are somewhat uncertain, we consider the following proposed parameter range: cloud mass (before any stars have formed) $M_{\rm{cl,0}} = 10^{5.5} - 10^{7.5}\,\Msun$, total stellar mass of the first generation of stars $M_{*,1} = 10^4-10^6\,\Msun$, total stellar mass of the second generation of stars (R136) $M_{*,2} = 2.2\times 10^4-10^5\,\Msun$, and cloud density $n = 10^2 - 10^{3.5}$\,cm$^{-3}$. For each combination $(M_{\rm{cl,0}}, M_{*,1}, M_{*,2}, n )$, we proceed as follows:

\begin{itemize}
\item At $t=0$ we place a star cluster of mass $M_{*,1}$ in the center of a cloud with constant density $n$ and mass $M_{\rm{cl,1}} = M_{\rm{cl,0}} - M_{*,1}$. This star cluster represents the older population in NGC~2070.
\item We use \textsc{warpfield} to model the evolution of the shell created by feedback from this star cluster in order to determine whether or not feedback successfully overcomes gravity.
\item If feedback is unable to overcome gravity, the cloud eventually recollapses. When the radius of the collapsing shell shrinks to 1\,pc we pause the simulation and define the time when this happens as the collapse time $t_{\rm{coll}}$. We reset the velocity of the shell to 0.
\item We assume that due to the cloud recollapse a second star cluster instantaneously forms, leading to renewed expansion. The age separation between the two star clusters is then $\Delta t_{\rm{age}} = t_{\rm{coll}}$. The second star cluster has a stellar mass of $M_{*,2} = 2.2\times 10^4\,\Msun$ or $1\times 10^5\,\Msun$ (representing the lower and upper mass estimates for R136). The expansion of the shell is now driven by a star cluster 
that consists of two distinct stellar populations. Before we continue the simulation we redistribute the shell mass into a spherical cloud with the same density as before, but with a smaller mass $M_{\rm{cl,2}} = M_{\rm{cl,1}} - M_{*,2}$.
\item At $t=8$\,Myr we stop the simulation.
\end{itemize}

\section{Results}

\begin{figure}
    \begin{center}
        \includegraphics[width=0.49\textwidth]{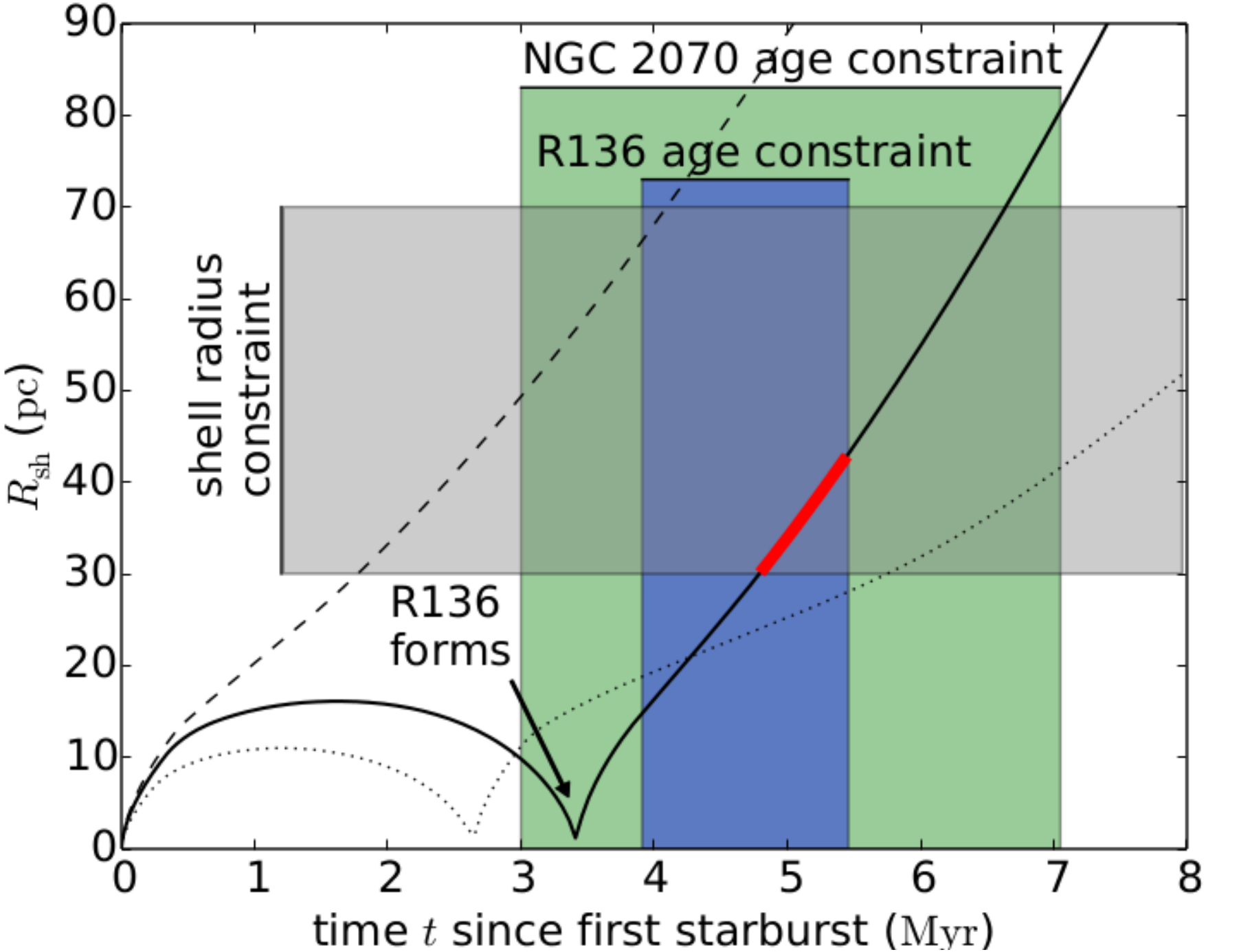} 
    \end{center}%
\caption{Time evolution of the shell radius (black solid line) for a model with $n = 2500$\,cm$^{-3}$, $M_{\rm{cl},0} = 10^6\,\Msun$, $M_{*,1} = 3\times 10^4\,\Msun$, $M_{*,2} = 10^5\,\Msun$. The various observational constraints are shown as gray, green and blue shaded areas, respectively. The thick red line shows where all three constraints are fulfilled. Also shown are examples of models which cannot fulfil all constraints: same parameters as before except for $M_{*,1} = 10^5\,\Msun$ (dashed line), $M_{*,1} = 2\times 10^4\,\Msun$ (dotted line)}
\label{fig:expansion_example}
\end{figure}

We will call a model (a given set of input parameters) consistent with the timeline described at the beginning of the previous section if at some time $t$ it can fulfil all of the criteria below:

\begin{enumerate}
\item \label{itm:age1} The older population in NGC~2070, which formed first, must have an age in the range $3 \leq t_{\rm age, 1} \leq 7$\,Myr;
\item \label{itm:age2} R136 forms in a second starburst and must have an age in the range $0.5 \leq t_{\rm{age},2} \leq 2$\,Myr;
\item\label{itm:Rsh} The shell must have a radius in the range $30 \leq \Rshell \leq 70$\,pc.
\end{enumerate}
A time at which all three criteria are met then marks the current, observed state of 30~Dor.
An example of a model which can reproduce all these constraints is shown in Figure~\ref{fig:expansion_example} (solid line), together with a model in which no second starburst occurs (and which is hence ruled out, dashed line) and a model in which feedback is too weak to push the shell to $30-70$\,pc in less than $2\,$Myr after R136 formed (dotted line).
%
%
\begin{figure*}
\centering
    \subfloat[Age separation between 1st and 2nd cluster]{%
        \includegraphics[width=0.5\textwidth]{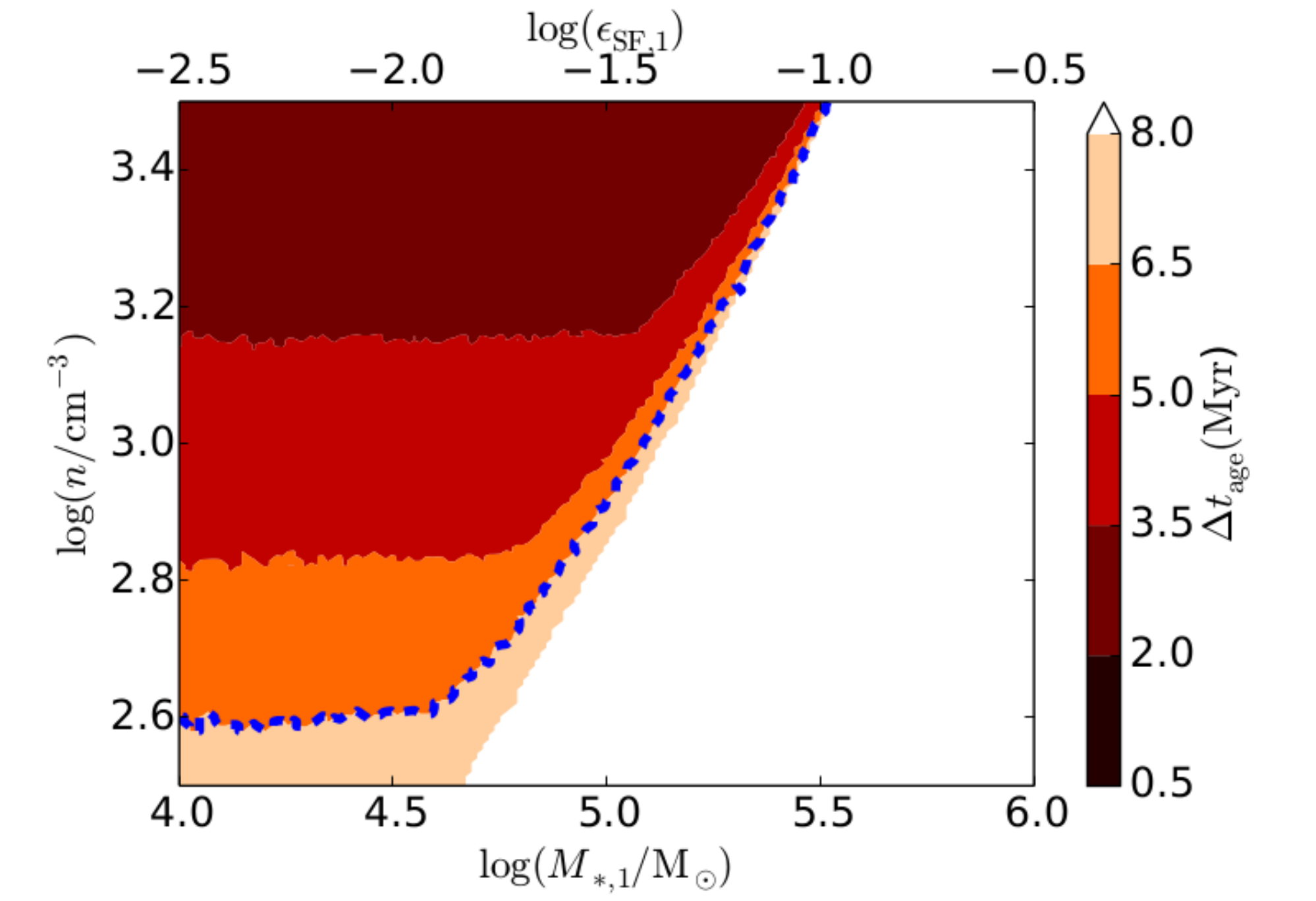}%
        \label{fig:age_separation}%
        }%
    \hfill%
    \subfloat[Current shell radius]{%
        \includegraphics[width=0.5\textwidth]{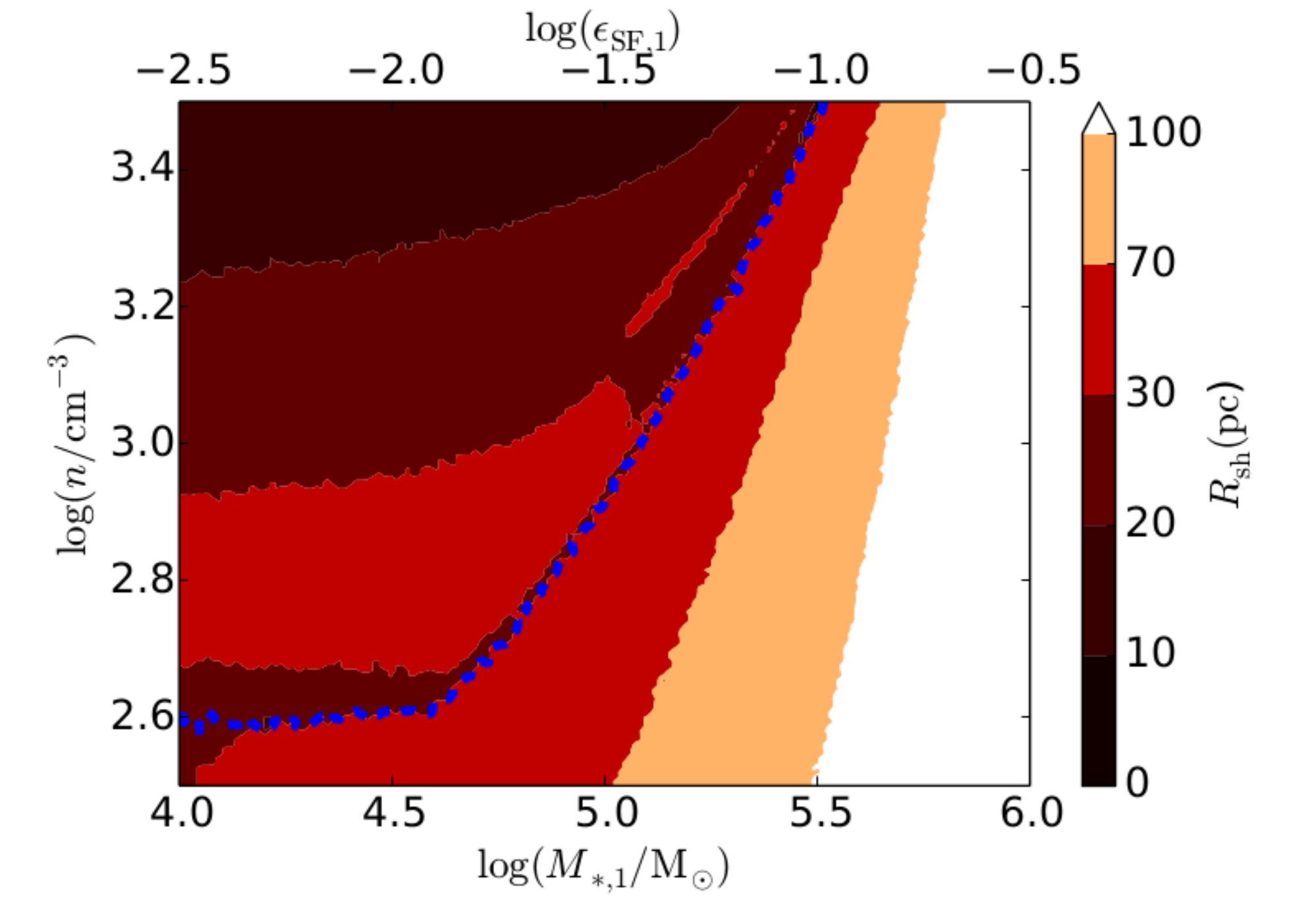}%
        \label{fig:Rsh}%
        }%
    \caption{(a) Age separation between the two clusters and (b) current shell radius, both plotted as a function of the mass $M_{*,1}$ of the first cluster and the cloud density $n$. The cloud mass before star formation $M_{\rm{cl},0} = 10^{6.5}\,\Msun$ and the mass of the second cluster $M_{*,2} = 10^5\,\Msun$ have been kept fixed. The blue dashed line divides regions with shells that recollapse on time scales shorter than 6.5\,Myr and are thus consistent with 30~Dor (above the line), and shells that take longer to recollapse (or never do).}
\end{figure*}
\begin{figure*}
\centering
    \subfloat[High mass R136 ($M_{*,2} = 10^5\,\Msun$)]{%
        \includegraphics[width=0.5\textwidth]{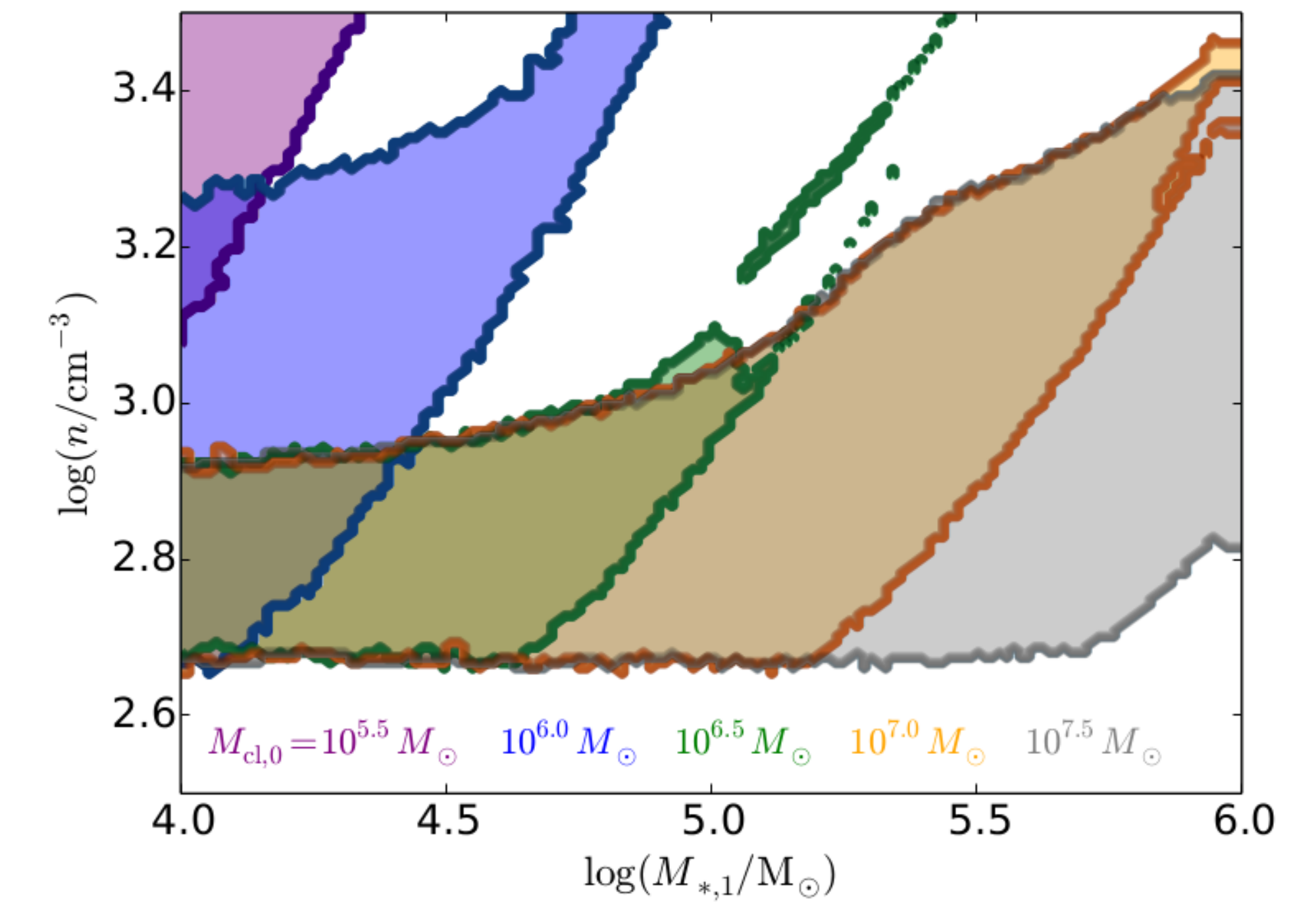}%
        \label{fig:valid_cloudvary_1e5}%
        }%
    \hfill%
    \subfloat[Low mass R136 ($M_{*,2} = 2.2\times 10^4\,\Msun$)]{%
        \includegraphics[width=0.5\textwidth]{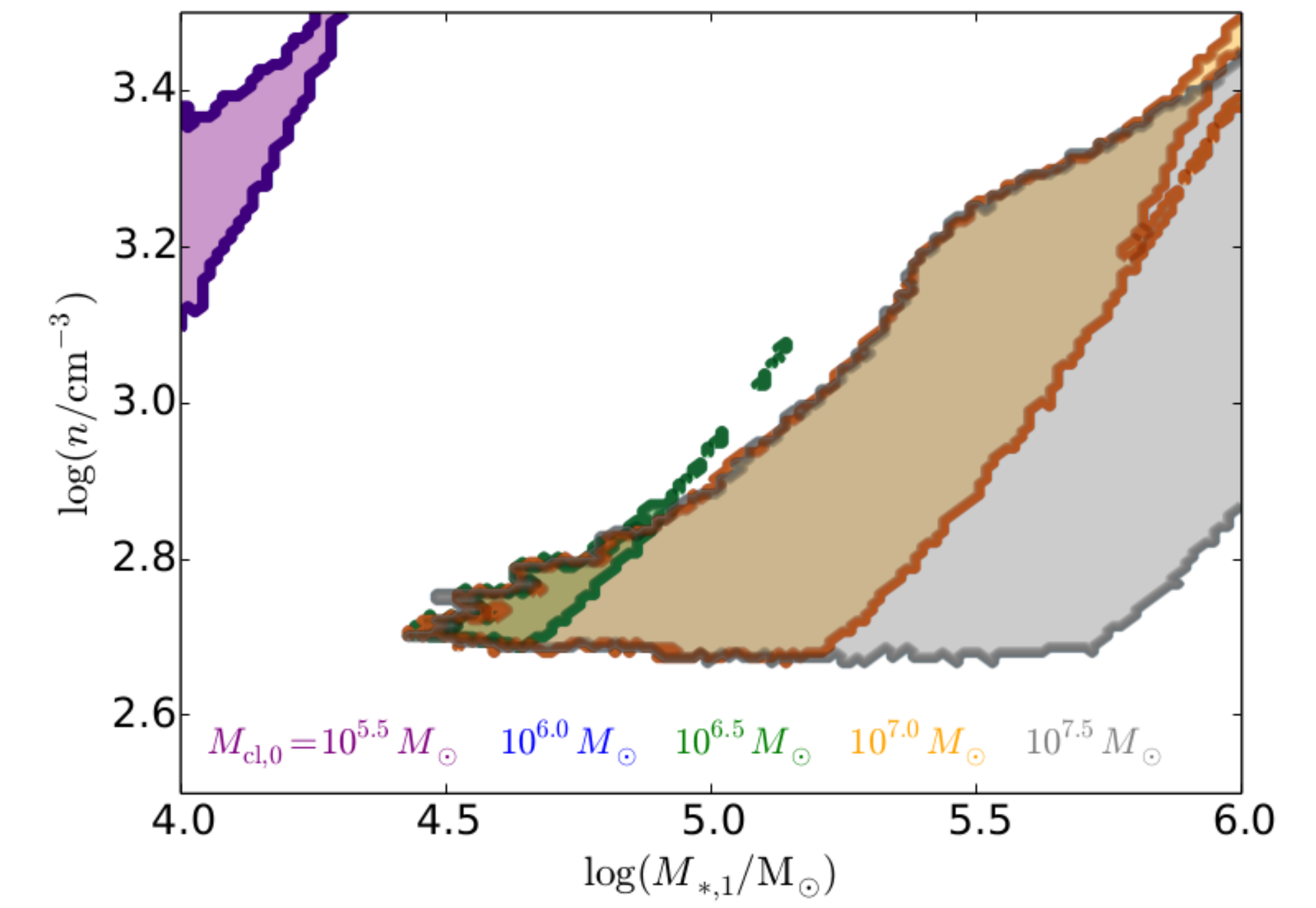}%
        \label{fig:valid_cloudvary_2e4}%
        }%
    \caption{Regimes of mass of the first cluster $M_{*,1}$ and cloud density $n$ for which models can reproduce both the observed current shell radius and the ages of the two clusters, shown for different initial cloud masses $\Mcc$ (colours). For an initial cloud mass $M_{\rm{cl,0}} = 10^6\,\Msun$ with $M_{*,2} = 2.2\times 10^4\,\Msun$ (blue in right figure) no model in the investigated parameter space can reproduce the observations.}
    \label{fig:valid_cloudvary}
\end{figure*}

A necessary (but insufficient) condition for meeting both constraints \ref{itm:age1} and \ref{itm:age2} is $1 \leq \Delta t_{\rm{age}}  \leq 6.5$\,Myr. 
In Figure~\ref{fig:age_separation} we show $\Delta t_{\rm{age}}$ for models with $M_{\rm{cl},0} = 10^{6.5}\,\Msun$ and $M_{*,2} = 10^5\,\Msun$. We see that we can already exclude low cloud densities and high masses for NGC~2070, as these do not produce the correct age separation. 
When $n \lesssim 10^{2.6}$\,cm$^{-3}$, it takes too long for the second cluster to form. Likewise, when $M_{*,1}\gtrsim 10^{5.5}\,\Msun$, corresponding to a star formation efficiency of the first star formation event $\epsilon_{\rm{SF},1} \equiv M_{*,1}/M_{\rm{cl,0}} \gtrsim 0.1$, the natal cloud is disrupted before a second starburst occurs ($\Delta t_{\rm{age}} = \infty$).

In Figure~\ref{fig:Rsh} the current shell radius is shown, determined in the following way:
First, for each model we determine the time span where condition \ref{itm:age1} and, if possible, also \ref{itm:age2} are met.\footnote{Models which cannot fulfil both \ref{itm:age1} and \ref{itm:age2} are already excluded; however, if they can reproduce the observed shell radius via a single, unstalled expansion of the shell they should be regarded as at least partly consistent.} We then select the radius (corresponding to a time in this time span) which is closest to the range $30-70$\,pc.
In Figure~\ref{fig:Rsh} we show that high density clouds cannot reproduce the observed current shell radius, even though they can form two clusters with the correct $\Delta t_{\rm{age}}$. Such models never simultaneously fullfil both \ref{itm:age2} and \ref{itm:Rsh}. Low densities are also excluded because the second starburst occurs later, with insufficient time left to fulfil both \ref{itm:age1} and \ref{itm:Rsh}.

Models which yield the correct ages and shell radius are fully consistent with the described picture of gravity-induced cloud recollapse, a second star formation event, and feedback-driven expansion of the shell to the current radius.
In Figure~\ref{fig:valid_cloudvary}, we summarize the allowed regime for various cloud and cluster properties. Independent of R136's mass, only clouds with average densities $n \gtrsim 10^{2.7}$\,cm$^{-3}$ can fulfil \ref{itm:age1}, \ref{itm:age2}, and \ref{itm:Rsh}. This threshold is also independent of cloud mass. 

In the model employed, clouds with a fixed density but different masses only differ in their size. We assume that parts of the cloud which have not been swept up by the cloud have enough turbulent support that they are stable against collapse. Hence, models with different cloud masses (but $n$, $\Mso$, $\Mst$ being equal) behave the same until the shell radius $\Rshell$ equals the initial cloud radius $\Rcl$. So if a model is consistent with observations and $\Rcl \geq 30$\,pc, models with larger cloud masses are also consistent.

Interestingly, if the mass of R136 is low (Figure \ref{fig:valid_cloudvary_2e4}), an initial cloud mass of $10^6\,\Msun$ is excluded: On the one hand, for a low-mass first generation (with weak stellar feedback), even the additional feedback of R136 is not enough to push the gas to 30\,pc or more. On the other hand, if the older generation is massive (and feedback is stronger), the whole cloud can get swept up by the shell, delaying its recollapse, and the second star formation event takes place at a time when the majority of the most massive stars have already died. As stars more massive than 40\,$\Msun$ do not contribute to SN feedback, again feedback is not strong enough to push gas far out. For $\Mcc > 10^6\,\Msun$, the cloud does not get swept up during the first expansion, the shell falls back earlier as it accumulates more mass, and some massive stars of the first generation can still contribute to the re-expansion. For $\Mcc < 10^6\,\Msun$, the second starburst turns a significant fraction of the available gas into stars (corresponding to a high star formation efficiency), so that it is much easier to push the remainder of the gas to 30\,pc or more.

\section{Discussion and Conclusion}

We have used the 1D stellar feedback scheme \textsc{warpfield} \citep{Rahner2017a}, which includes mechanical and radiative feedback as well as gravity, to test whether the properties of NGC~2070 and the ISM around it may be a result of initial gas expulsion from the cluster, followed by a gravity induced cloud recollapse which leads to a second star formation event, creating the dense cluster R136, and finally renewed shell expansion.
We show that there is a reasonable parameter regime in which both the observed ages of the young and old stellar population in NGC~2070 and the size of the current feedback-driven shell around NGC~2070 can be reproduced.

If the parental cloud does not turn more than $\sim 10\,$\% of the available gas into stars in a very short period of time and if the density of the cloud is quite high ($\gtrsim 500$\,cm$^{-3}$), due to shell recollapse a second star formation event \textit{at the same location} several Myr later has to be expected and no further mechanisms like capture of field stars or gas retention inside the cluster need be invoked. 
Typical star formation efficiencies of nearby molecular clouds are less than 10\,\% \citep{Lada2010}, but only the densest GMCs in the LMC achieve densities of $\sim$500\,cm$^{-3}$ \citep{Hughes2010}.
If we settle for $n = 500$\,cm$^{-3}$ and an initial cloud mass of $\Mcc = 3\times 10^6\,\Msun$, we find a total stellar mass of NGC~2070 in the range $5\times 10^4\,\Msun \leq \Mso + \Mst \leq 1.5\times 10^5\,\Msun$ which agrees well with mass estimates from observations. 

Many more young star clusters could host multiple generations resulting from ISM recollapse, but we may not yet be able to identify them as the window of opportunity for unambiguously identifying such events is short. Within the local group 30~Dor is the only known system that is simultaneously: 1) massive enough for feedback to initially fail, 2) in close enough proximity and with sufficiently low extinction for photometry and spectral observations to distinguish superimposed compact clusters, and 3) sufficiently young, making multiple generations of star formation with a few Myr offset distinct in observations. We speculate James Webb Space Telescope will reveal a population of post-recollapse, highly embedded clusters undergoing an unexpected and intense second event of star formation.

\section*{Acknowledgements}

We acknowledge funding from the Deutsche Forschungsgemeinschaft in the Collaborative Research Centre (SFB 881) ``The Milky Way System'' (subprojects B1, B2, and B8), the Priority Program SPP 1573 ``Physics of the Interstellar Medium'' (grant numbers KL 1358/18.1, KL 1358/19.2, and GL 668/2-1), and the European Research Council in the ERC Advanced Grant STARLIGHT (project no. 339177).

\begin{footnotesize}
\bibliographystyle{mn2e}  
\bibliography{mylibrary}
\end{footnotesize}

\bsp    
\label{lastpage}
\end{document}